\documentclass[aps,twocolumn,superscriptaddress,amsmath,longbibliography]{revtex4-2}  
\usepackage{graphicx}  
\usepackage{dcolumn}   
\usepackage{bm}        
\usepackage{amsmath,amsfonts,amssymb}   
\usepackage[colorlinks]{hyperref}
\usepackage{floatrow}
\usepackage[caption=false]{subfig}
\usepackage{xcolor}

\newtheorem{theorem}{Theorem}

\newcommand{\CC}{\mathbb{C}}
\newcommand{\RR}{\mathbb{R}}

\newcommand{\ConditionallyIndependent}[3]{#1 \perp\kern-5pt \perp #2 \mid #3}

\newcommand{\set}[1]{\lbrace #1 \rbrace}

\newcommand{\norm}[1]{\lVert #1 \rVert}

\newcommand{\ket}[1]{\lvert #1 \rangle}

\DeclareMathOperator{\sinc}{sinc}

\newcommand{\beq}{\begin{equation}}
\newcommand{\eeq}{\end{equation}}

\begin{document}

\title{Compressed Qubit Noise Spectroscopy:\\ Piecewise-Linear Modeling and Rademacher Measurements}
\author{Kaixin Huang}
\email{kxhuang@umd.edu}
\affiliation{Joint Quantum Institute (JQI), NIST/University of Maryland, College Park, MD 20742, USA}
\affiliation{Joint Center for Quantum Information and Computer Science (QuICS),
NIST/University of Maryland, College Park, MD 20742, USA}
\author{Demitry Farfurnik}
\affiliation{Department of Electrical and Computer Engineering, North Carolina State University, Raleigh, NC, 27695, USA}
\affiliation{Department of Physics, North Carolina State University, Raleigh, NC, 27695, USA}
\author{Dror Baron}
\affiliation{Department of Electrical and Computer Engineering, North Carolina State University, Raleigh, NC, 27695, USA}
\author{Yi-Kai Liu}
\affiliation{Joint Center for Quantum Information and Computer Science (QuICS),
NIST/University of Maryland, College Park, MD 20742, USA}
\affiliation{Applied and Computational Mathematics Division, National Institute of Standards and Technology (NIST), Gaithersburg, MD 20899, USA}

\noaffiliation       

\date{\today}

\begin{abstract}
Random pulse sequences are a powerful method for qubit noise spectroscopy, enabling efficient reconstruction of sparse noise spectra. Here, we advance this method in two complementary directions. First, we extend the method using a regularizer based on the total generalized variation (TGV) norm, in order to reconstruct a larger class of noise spectra, namely \textit{piecewise-linear} noise spectra, which more realistically model many physical systems. We show through numerical simulations that the new method resolves finer spectral features, while maintaining an order-of-magnitude speedup over conventional approaches to noise spectroscopy. Second, we simplify the experimental implementation of the method, by introducing \textit{Rademacher measurements} for reconstructing sparse noise spectra. These measurements use pseudorandom pulse sequences that can be generated in real time from a short random seed, reducing experimental complexity without compromising reconstruction accuracy. Together, these developments broaden the reach of random pulse sequences for accurate and efficient noise characterization in realistic quantum systems.
\end{abstract}

\maketitle

\section{Introduction}

Characterizing the spectral properties of environmental noise is essential for understanding and mitigating decoherence in quantum systems. In particular, the accurate reconstruction of the noise spectral density, \( S(\omega) \), informs both noise-resilient control protocols and error mitigation strategies in quantum technologies~\cite{Cywinski2008, Biercuk2011, Degen2017}. Dynamical decoupling (DD) techniques enable indirect access to \( S(\omega) \) by modulating a system's sensitivity to environmental fluctuations via tailored control sequences. However, conventional methods of DD often rely on coarse frequency resolution or assumptions of spectral smoothness, limiting their effectiveness in resolving structured and non-smooth features~\cite{bylander2011noise, alvarez2011measuring, multiseq, norris2016qubit, chan2018assessment}.

To improve the resolution and resource efficiency of noise spectroscopy, \textit{compressed sensing} (CS) has emerged as a powerful framework for spectral reconstruction in quantum noise spectroscopy, particularly when the spectrum possesses sparse structure in a suitable representation~\cite{Norris2016, Roger2018}. Recent work has introduced \textit{random pulse sequences} as a tool for the direct measurement of arbitrary linear functionals of $S(\omega)$, enabling the CS reconstruction of sharp spectral peaks with 
fewer measurements~\cite{huang2025random}. 

While this approach is effective for perfectly sparse spectra, it leaves two key challenges unresolved: can this method be applied to a larger class of noise spectra that are not perfectly sparse? And can the practical difficulty of applying the desired random pulse sequences in an experiment be reduced? This paper addresses both of these challenges separately.

First, many realistic noise processes, including charge and magnetic noise in solid-state qubits, exhibit more complicated spectral features rather than sparse peaks represented by simple delta functions~\cite{stockill2016quantum, farfurnik2021all}.
For example, the spectra of quantum dot systems often combine narrow resonances with broad, slowly varying backgrounds. This motivates the need for models that retain sparsity while capturing richer structures. \textit{Piecewise-linear}, and more generally piecewise-polynomial, functions provide such a representation.

In this work, we develop a CS protocol for reconstructing piecewise-linear noise spectra by enforcing sparsity on the second derivative \( S''(\omega) \). Specifically, we apply an \( L_1 \)-norm regularization on \( S''(\omega) \), which promotes reconstructions that are globally continuous but exhibit sparse curvatures. This approach aligns with recent advances in  \textit{total generalized variation} (TGV)~\cite{bredies2010total, knoll2011second, karahanoglu2011signal} and higher-order sparse regularization techniques for inverse problems~\cite{Figueiredo2006, Candes2008}. It can also be viewed as an extension of well-known work on compressed sensing in the context of medical imaging, for piecewise-constant images in two dimensions, based on regularization using the total variation (TV) norm \cite{needell2013near, poon2015role, krahmer2018total, genzel2022compressed}. 

Our method extends this to the problem of compressed sensing of a piecewise-linear function using TGV regularization, in one dimension. This particular setting appears to be more challenging from a theoretical perspective, and beyond the scope of the above works. Nonetheless, our numerical simulations on simulated spectra with varying degrees of piecewise-linearity demonstrate high-fidelity reconstructions.

Second, we consider the experimental implementation of random pulse sequences, which can be resource-intensive. In particular, previous work proposed the use of random pulse sequences with complex, long-range correlations, which are difficult to generate in real time. Instead, the entire pulse sequence typically had to be generated ``offline,'' using a desktop computer, then transferred to an arbitrary waveform generator (AWG), before being applied to a physical qubit. The loading and storage of a great number of different pulse sequences creates a bottleneck in experiments. 

To address this bottleneck, we introduce a simplified variant of the method, using uncorrelated random pulses, which we call \textit{Rademacher measurements}. These pulse sequences are easier to generate in real time, using a short random seed and a high-quality pseudorandom number generator \footnote{The output of a cryptographic pseudorandom number generator is almost certainly good enough for this purpose.}, so that the entire pulse sequence never has to be stored in memory. This opens up the possibility of generating the random pulse sequences ``on the fly'' using a field-programmable gate array (FPGA), without any need for an AWG.

In this scheme, the total evolution time is divided into equal segments indexed by $m=1,2,\ldots,M$, each time segment is assigned a Rademacher random variable $U_m$ (i.e., an independent random variable that takes on the values 1 and $-1$ with equal probability), and rotation $\pi$-pulses are applied to the qubit whenever $U_{m+1} \neq U_m$. 

At first glance, it may seem that these pulse sequences consist of nothing more than white noise, and thus can only measure a single degree of freedom of the environment of the qubit. We get around this difficulty by using a simple trick: we perform many measurements using the \textit{same} realization of the random variables $U_m$ (i.e., by running a pseudorandom number generator many times using the same seed). This lets us measure the signal associated with the \textit{fluctuations} in a single realization of the pulse sequence. By repeating this measurement with different fluctuations, we collect enough information to fully characterize the environment of the qubit.

Despite their simplicity, we show via theory and simulations that Rademacher measurements retain the same compressed-sensing efficiency as Fourier-based random pulses~\cite{huang2025random}, while substantially reducing experimental complexity. Interestingly, these Rademacher measurements are similar in spirit to Gaussian measurements in compressed sensing, which are well-studied \cite{candes2008introduction, foucart2013mathematical}. But our Rademacher measurements have an unusual feature, which is different from these older works on compressed sensing: our measurements depend quadratically (rather than linearly) on the random variables $U_m$, as shown in Eq.~(\ref{eqn-lrm}). For this reason, our recovery guarantee for Rademacher measurements relies on more recently-developed techniques, involving the recovery of structured low-rank matrices \cite{huang2025low, kueng2017low}.

Finally, we extend our investigation to realistic physical systems, focusing on the noise spectra of self-assembled quantum dots \cite{farfurnik2021all}. We demonstrate the efficiency of the TGV-based CS method, combine it with the Rademacher measurements technique, and demonstrate its potential for rapidly reconstructing realistic noise spectra. We also explore modifications of the protocol that further reduce the number of control pulses without significant loss in reconstruction accuracy. These results broaden the applicability of random-pulse-based noise spectroscopy and improve its practicality for near-term quantum devices.

The remainder of the paper is organized as follows. The basic model and background are introduced in Sec.~\ref{model}. The two main techniques of the paper are developed in detail in Secs.~\ref{piecewise-th} and~\ref{unco_th}. We explore applications of our techniques on realistic systems in Sec.~\ref{appli}.
We conclude
with a summary and a discussion of open problems in Sec.~\ref{out}.

\section{\label{model}Noise Spectroscopy Model}

We consider a single qubit (``the system") subject to pure dephasing due to coupling with a classical fluctuating environment. The Hamiltonian is given by
\beq
\hat{H}(t) = \hat{H}_0 + \hat{H}_V(t) = [\Omega + V(t)] \sigma_z,
\eeq
where $\hat{H}_0 = \Omega \sigma_z$ is the system Hamiltonian, and $\hat{H}_V(t) = V(t) \sigma_z$ describes the coupling of the spin qubit to a stochastic noise process $V(t)$ from the bath (e.g., fluctuations of a magnetic field). The noise spectrum $S(\omega)$ is defined as the Fourier transform of the autocorrelation function $g(t - t')=\langle V(t) V(t') \rangle_V$, where $\langle \cdot \rangle_V$ denotes the average with respect to the ensemble of $V(t)$~\cite{theoryofoqs, qds}. We further assume that the noise spectrum is band-limited, i.e., $S(\omega) = 0$ for $|\omega| > \omega_c$, where $\omega_c$ is a high-frequency cutoff \cite{viola1999dynamical, khodjasteh2011limits}.

To estimate the spectrum $S(\omega)$, we employ the \textit{filter function formalism} \cite{paz2014general}, which applies control pulses to change the frequency domain susceptibility of the qubit to noise. A control sequence can be associated with a filter function, $F(\omega)$, that characterizes the qubit's spectral response. For a qubit initialized and measured in the $\ket{+}=\frac{1}{\sqrt2}(\ket{0}+\ket{1})$ state, the survival probability over evolution time $T$ can be approximated as
\begin{equation}
\label{prob}
    P(T) \approx \frac{1}{2} + \frac{1}{2} \exp\left(-\int_{-\omega_c}^{\omega_c} d\omega\, S(\omega) F(\omega) \right).
\end{equation}
The decay exponent, $\chi(T) = \int d\omega\, S(\omega) F(\omega)$, quantifies the extent to which the noise spectrum overlaps with the control's frequency response. Note that $F(\omega)$ depend implicitly on the evolution time $T$.

Equation (\ref{prob}) provides a pathway for estimating the spectrum of a noisy environment via linear inversion. We apply a set of $m$ control sequences on the system,  then measure the corresponding survival probabilities $P_k$, $k = 1, \ldots, K$. For each $k$, one obtains estimates of $\chi_k = -\log(2P_k - 1)$. To make the inversion tractable, we discretize the frequency axis at $N$ points $\{\omega_n\}$, leading to the approximation
\begin{equation}
\label{chiform}
    \chi_k = \sum_{n=1}^{N} F_{k}(\omega_n) S(\omega_n)+\epsilon_k.
\end{equation}
Here $\epsilon_k$ represents the error in the $k$-th measurement. In matrix form, Eq.~(\ref{chiform}) becomes
\begin{equation}
\label{matr_1}
    \boldsymbol{\chi} = \mathbf{F} \mathbf{S}+\boldsymbol{\epsilon},
\end{equation}
where $\boldsymbol{\chi} \in \mathbb{R}^K$ contains the measured decay exponents, $\mathbf{F} \in \mathbb{R}^{K \times N}$ encodes the filter functions, and $\mathbf{S} \in \mathbb{R}^N$ represents the discretized noise spectrum. The estimation of $\mathbf{S}$ then reduces to solving a linear regression problem. 

A common way to solve this problem is to apply the Carr-Purcell-Meiboom-Gill (CPMG) pulse sequence with evenly spaced rotation $\pi$ pulses, whose filter function is approximately the Dirac $\delta$-function. One can then solve Eq.~(\ref{matr_1}) via nonnegative least squares (NNLS) or other deconvolution methods \cite{huang2025random}. Intuitively, each CPMG sequence captures a narrow range of the spectrum, and one needs to apply many sequences to cover all frequency ranges.  As a result, the number of measurements needed, $K$, is of the same order as the discrete number in the frequency domain, $N$. However, such numerous measurements are not always necessary for extracting noise spectra with some prior knowledge. For example, we may know \textit{a priori} that the spectrum we are investigating has some sparse structure in a suitable representation~\cite{Norris2016, Roger2018}. 

Recent studies have shown that random pulse sequences \cite{huang2025random} and compressed sensing \cite{candes2008introduction, candes2011probabilistic} can be used to greatly speed up the reconstruction of sparse spectra. By generating random pulse sequences whose filter functions are Fourier basis functions, the number of measurements needed, $K$, is reduced from $O(N)$ to $O(s \log N)$, where $s$ is the sparsity of $\mathbf{S}$. 

However, the previous approaches are only capable of fully reconstructing ideal sparse spectra with Dirac $\delta$ function shapes. For realistic spectra with more complicated shapes,  one could only detect the centers of the peaks, and not a complete characterization of their shapes~\cite{huang2025random}. In addition, for each measurement of the decay exponent, we need to generate a set of different random pulse sequences and load them into the experimental control system. This loading procedure could be time-consuming on current quantum 
platforms.

To overcome the above limitations, we will advance this method in two complementary directions. We first extend the method to reconstruct piecewise-linear spectra, and demonstrate the efficiency of the method through numerical simulations. Next, we introduce a simplified variant of the method using Rademacher measurements, which greatly reduces experimental complexity for reconstructing sparse spectra.

\section{\label{piecewise-th}Estimation of Piecewise-Linear Spectra}
In this section, we show how random pulse sequences can be adapted to estimate piecewise-linear noise spectra. Following \cite{huang2025random}, we generate random pulse sequences with total time $T$ and $M$ equal segments, such that $T = M\tau$. Here $\tau$ is the minimum time between two consecutive pulses in the sequence, and we choose $\tau < 1/\omega_c$, where $\omega_c$ is the high-frequency cutoff of the noise spectrum $S(\omega)$. The filter functions of those sequences can be set to approximate the Fourier basis functions,
\begin{equation}
F_k(\omega_n) \propto \cos (j_k \omega_n \tau).
\end{equation}
Here, $j_k$ is a random integer chosen between $[0, M]$.
If we assume that $S(\omega_n)$ is sparse with at most $s$ nonzero points, the discretized spectrum $\mathbf{S^*}:\: G_N \rightarrow \RR$ (where $G_N$ is the set of grid points) can then be recovered by solving a convex optimization problem,
\begin{equation}
\label{CSconvec-0}
\begin{aligned}
\mathbf{S}^* 
    &= \underset{\mathbf{S}: G_N \to \mathbb{R}}{\arg\min} 
       \left( \, \| \boldsymbol{\chi} - \mathbf{F}\mathbf{S} \|_{L_2}^2 
       + \lambda \, \|\mathbf{S}\|_{L_1} \, \right),
\end{aligned}
\end{equation}
where $\lambda$ can be optimized with cross-validation. The solution $\mathbf{S^*}$ to Eq.~(\ref{CSconvec-0}) is an accurate estimation of $\mathbf{S}$ if the number of generated Fourier functions satisfies $K \geq {\Omega}(s\log N)$.

We then generalize the result to noise spectra that are piecewise linear. This class of functions appears in various physical systems, where the underlying noise processes vary smoothly within frequency bands, but exhibit kinks or discontinuities at a few locations. A prototypical example includes Lorentzian kinks with $1/f$ noise background regularized at high and low frequencies, or engineered environments with structured frequency response \cite{schultz2021schwarma, murphy2022universal}. The classification of a spectrum as piecewise linear is useful because the \textit{second-order derivative} is still sparse, which allows us to perform CS.

We define the second-order discrete difference operator $\mathbf{D}^2 \in \mathbb{R}^{(N-2)\times N}$ as
\begin{equation}
(\mathbf{D}^2 \mathbf{S})_n = S_{n+2} - 2S_{n+1} + S_n, \quad 1 \leq n \leq N-2.
\end{equation}
This operator acts as a discrete Laplacian and evaluates local curvature; for a piecewise linear function, $\mathbf{D}^2 \mathbf{S}$ is sparse, with nonzero entries only at the discontinuity points of the slope. Note that the cosine functions are the eigenfunctions of the discrete Laplacian operator, 
hence
\begin{equation}
    \sum_{n=1}^{N} F_{k}(\omega_n) S(\omega_n)\approx \frac{1}{(j_k\tau)^2}\sum_{n=1}^{N} F_{k}(\omega_n) \mathbf{D}^2S(\omega_n).
\end{equation}

We define $\Delta = \mathbf{D}^2 \mathbf{S}$ as the second order derivatives of the spectrum. As a result, similar to problem (\ref{CSconvec-0}), we are guaranteed to recover $\Delta$ by solving the following optimization problem:
\begin{equation}
\label{CSconvec-1}
    \begin{aligned}
\Delta^* 
    &= \underset{\Delta: G_N \to \mathbb{R}}{\arg\min} 
       \left( \, \| \boldsymbol{\chi^*} - \mathbf{F}\Delta \|_{L_2}^2 
       + \lambda \, \|\Delta\|_{L_1} \, \right),
\end{aligned}
\end{equation}
where $\boldsymbol{\chi^*} = ((j_1\tau)^2\chi_1, \ldots, (j_K\tau)^2\chi_K)$. We define $s^*$ as the sparsity in $\Delta$. Intuitively, $s^*$ is the number of spectral kinks.  The solution $\Delta^*$ to (\ref{CSconvec-1}) is exact when the number of generated Fourier functions satisfies $K \geq {\Omega}(s^*\log N)$. One may then reconstruct \(\mathbf{S}\) by double integration (solving \(\mathbf{D}^2\mathbf{S}=\Delta^*\)) with boundary constraints (e.g., \(S_1=S_N=0\), or by low-frequency anchors). 

However, the above procedure can be numerically unstable: it can amplify low-frequency noise, and accumulate error when integrating $\mathbf{D}^2\mathbf{S}^*$. In practice, we can solve a one-step convex problem that directly estimates 
 $\mathbf{S}$ while promoting second-order sparsity:
\begin{equation}
\label{CS_f}
    \begin{aligned}
\mathbf{S}^* 
    &= \underset{\mathbf{S}: G_N \to \mathbb{R}}{\arg\min} 
       \left( \, \| \boldsymbol{\chi} - \mathbf{F}\mathbf{S} \|_{L_2}^2 
       + \lambda \, \|\mathbf{D}^2\mathbf{S}\|_{L_1} \, \right).
    \end{aligned}
\end{equation}
Equation (\ref{CS_f}) can also be viewed as a second-order TGV-regularized~\cite{bredies2010total} optimization problem. 

While first-order TGV with Fourier measurements has a well-developed theory with rigorous error bounds~\cite{needell2013near,poon2015role, candes2011compressed, ongie2016off, krahmer2018total, genzel2022compressed}, the second-order case lacks equally general bounds in our setting. However, studies have shown that the second order TGV can be applied for signal reconstruction with undersampled discrete Fourier measurements, and yields results that are superior to conventional methods~\cite{knoll2011second, liu2015fast}. In this work, we show through simulations on piecewise linear spectra that our problem (\ref{CS_f}) can be solved efficiently, and produces accurate reconstructions of $\mathbf{S}$.

\begin{figure}
\begin{centering}
\sidesubfloat[]{\label{four_illus}%
  \includegraphics[width=0.8\textwidth]{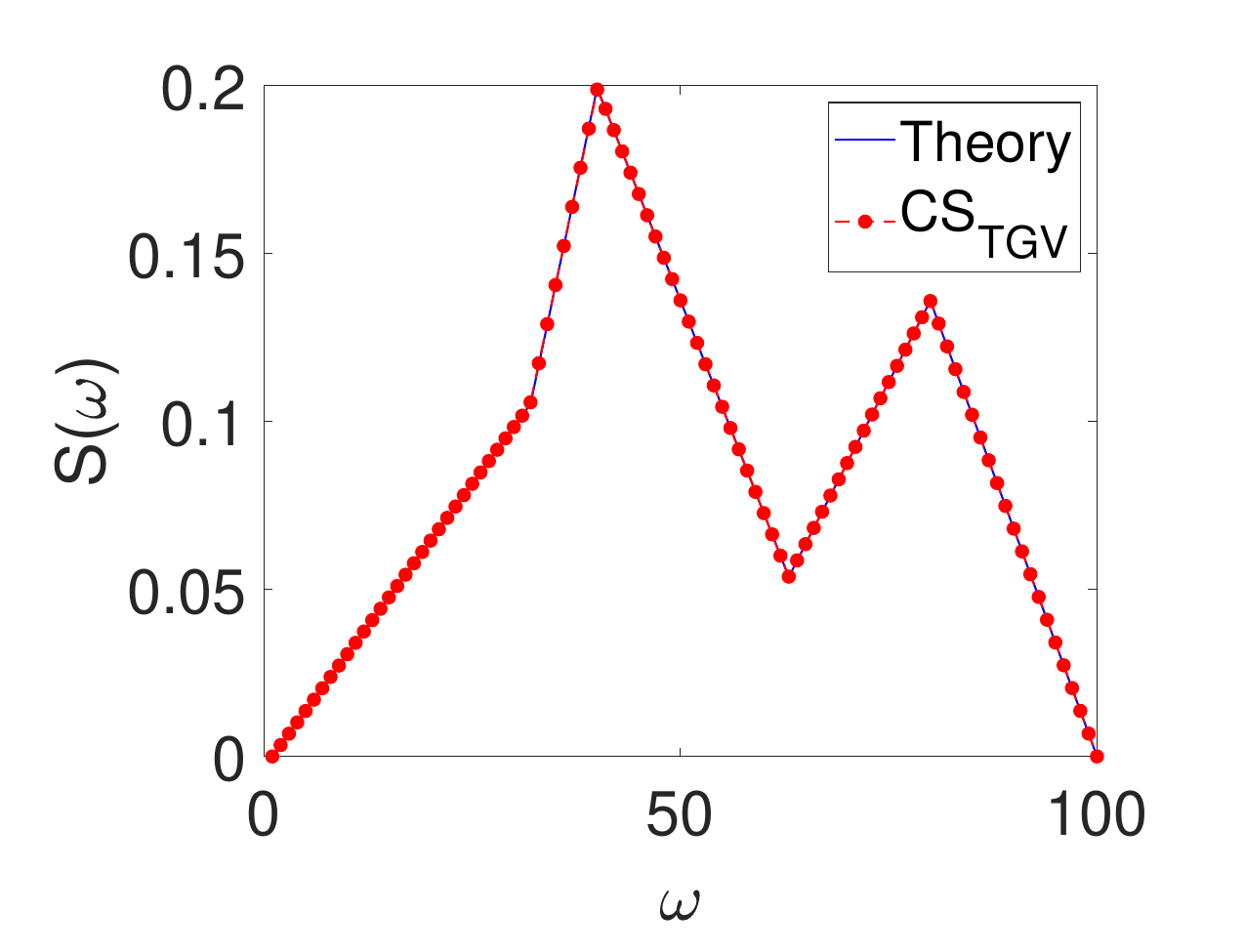}%
}
\par\end{centering}
\begin{centering}
\sidesubfloat[]{\label{four_phase}%
  \includegraphics[width=0.8\textwidth]{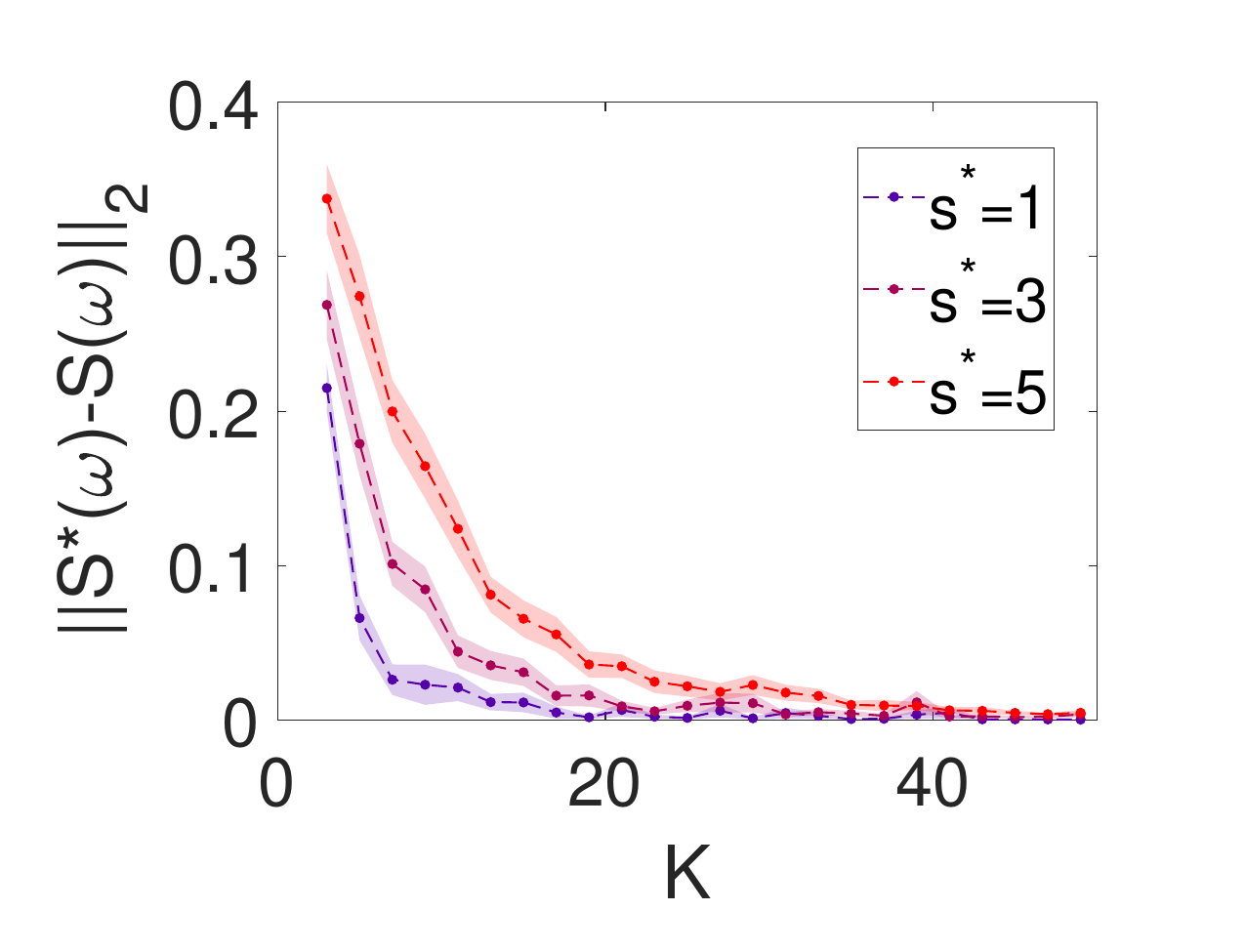}%
}
\par\end{centering}
\caption{
(a) A reconstruction of an ideal piecewise-linear spectrum using the CS$_\text{TGV}$ method. The solid line represents a randomly generated spectrum with $N=100$ grid points. The second-order derivative of this spectrum is 4-sparse. The red dots represent the reconstructed spectrum using CS$_\text{TGV}$ based on $K=20$ different Fourier basis functions. For each Fourier basis function, we generate random pulse sequences with $(M, N_1, N_2)=(100, 100, 50)$. (b) The accuracy of CS$_\text{TGV}$ ($(M, N_1, N_2)=(100, 100, 50)$) in reconstructing ideal spectra as a function of the number of Fourier basis functions. Different curves represent different sparsities $s$ of the second-order derivatives, considering  40  randomly generated spectra with $N=100$, normalized so that the $L_2$ norm equals $1$. Each simulation is repeated 100 times and the shaded areas represent the 95\% confidence regime. 
}
\label{rpsillus}
\end{figure}

Fig.~\ref{four_illus} shows a numerical example demonstrating the effectiveness of CS with second-order TGV penalty (CS$_\text{TGV}$) in reconstructing an ideal piecewise-linear noise spectrum. The true spectrum, which is 4-sparse in the second-order difference domain over \( N = 100 \) grid points, is represented as a solid blue curve. The spectrum recovered from CS$_\text{TGV}$ is shown as red circles and is obtained using only \( K = 20 \) random Fourier basis measurements. For each Fourier basis, we generate $N_1=100$ different random pulse sequences with each individual sequence repeated $N_2=50$ times.
The reconstruction closely aligns with the original signal, confirming that CS$_\text{TGV}$ is capable of accurately recovering the spectrum structure with far fewer samples than grid points.

To further investigate the performance of CS$_\text{TGV}$, we assess its reconstruction accuracy across various levels of second-order sparsity, \( s^* \), as a function of the number of Fourier basis measurements, \( K \); see Fig.~\ref{four_phase}. For each sparsity (different dotted lines in Fig.~\ref{four_phase}), we generate 100 random piecewise linear spectra with fixed norm, each with \( N = 100 \) grid points, to obtain the averaged accuracy. The accuracy is quantified as the \( L_2 \) norm of the difference between the discrete true spectrum, \( S(\omega) \), and the CS estimate \( S^*(\omega) \). (In our experiments, after choosing the units of $\omega$ appropriately, the grid points are consecutive integers, so the $L_2$ norm can be approximated by a sum over grid points, $\norm{F}_{L2} \approx (\sum_j |F(j)|^2)^{1/2}$.) For each sparsity, the accuracy increases sharply as $K$ increases, revealing the potential of CS$_\text{TGV}$ to reconstruct piecewise linear spectra.

\section{\label{unco_th}Rademacher Measurements for Sparse Spectra}
In the previous section, we described improvements to the numerical algorithms for reconstructing sparse and piecewise-linear noise spectra. Now we turn our attention to different classes of random pulse sequences that can be used for CS. In this section, we will demonstrate a new way of reconstructing sparse noise spectra using Rademacher random pulse sequences and CS. 

This new class of random pulse sequences is much simpler compared to the previous work \cite{huang2025random}, as it only contains uncorrelated $\pi$-pulses. Specifically, we assign a sequence of Rademacher random variables (sampled independently and uniformly from $\set{1,-1}$), denoted $\vec{U}  = (U_1,\ldots,U_M) \in \set{1,-1}^M$, to the $M$ segments of the total experimental time $T$. In other words, the probability that $U_m = 1$ is $p = 0.5$ and so is the probability that we generate a $\pi$-pulse in between the time segments. (In the same way as before, we write $T = M\tau$, and we assume $\tau < 1/\omega_c$, where $\omega_c$ is the high-frequency cutoff of the noise spectrum $S(\omega)$.)

Differing from the previous approach, the random pulses are completely uncorrelated, and the Rademacher measurements do not involve averaging the filter function over many independent realizations of $\vec{U}$, but rather focus on the specific filter function of a \textit{single} realization of $\vec{U}$. (Note that the \textit{same} realization of $\vec{U}$ can be generated many times, by using a pseudorandom number generator with the same random seed.) In other words, rather than taking an average of many filter functions in order to ``smooth out'' their random fluctuations, we take a single filter function, and we use its large fluctuations to sense the noise spectrum.

This filter function can be written as
\begin{equation}
\begin{split}
F_k(\omega) 
&= \tau^2 \sinc^2(\tfrac{\omega\tau}{2}) \Bigl| \sum_{m=1}^M U_m e^{i\omega m\tau} \Bigr|^2 \\
&= \tau^2 \sinc^2(\tfrac{\omega\tau}{2}) \vec{U}^T A(\omega) \vec{U}, 
\end{split}
\end{equation}
where we define the matrix $A(\omega) \in \CC^{M\times M}$ whose $(m,m')$ entry is 
\begin{equation}
A_{m,m'}(\omega) = e^{i\omega (m-m')\tau}.
\end{equation}

We can further define the matrix $B(S) \in \RR^{M\times M}$ whose $(m,m')$ entry is 
\begin{equation}\label{bdef}
B_{m,m'}(S) = \hat{S}_{m-m'} \equiv \int_{-\pi/\tau}^{\pi/\tau} d\omega S(\omega)\sinc^2(\tfrac{\omega\tau}{2}) A_{m,m'}(\omega),
\end{equation}
where $\hat{S}_{m-m'}$ is equal to the $(m-m')$'th coefficient of the Fourier series expansion of $S(\omega)\sinc^2(\tfrac{\omega\tau}{2})$.  We find that $B(S)$ is a Toeplitz matrix, in which each descending diagonal from left to right is constant. The eigenvalues of $B(S)$ equal $M\mathbf{S'}: G_M \rightarrow \mathbb{R}$, which is the discretized spectrum of $MS(\omega)\sinc^2(\tfrac{\omega\tau}{2})$ on $M$ grid points. 

We can  now rewrite Eq.~(\ref{matr_1}) as a linear operator $\boldsymbol{A}(B):\RR^{M\times M} \rightarrow \RR^K$ that maps $B(S) \in \RR^{M\times M}$ to $\boldsymbol{\chi} \in \RR^{K}$:
\begin{equation}
\label{eqn-lrm}
\boldsymbol{\chi} = \boldsymbol{A}(B(S)) = \frac{\tau^2}{2\pi} \mathbf{U}^T B(S) \mathbf{U}+\boldsymbol{\epsilon}.
\end{equation}
Here, $\mathbf{U} \in \{1, -1\}^{K \times M}$ encodes $K$ sets of Rademacher sequences. 

 Similarly to the previous CS scenario, if we assume that $\mathbf{S'}$ is sparse with sparsity $s$, the Toeplitz matrix $B(S)$ is also of low rank $s$. Thus, the estimation of $\mathbf{S'}$  is equivalent to the recovery of the low-rank matrix $B(S)$: 
\begin{equation}
    \label{prb-lrm}
        \begin{aligned}
   B^* 
    = \underset{B\in \mathbb{R}^{M\times M}, B \text{ is Toeplitz}}{\arg\min} 
       \left( \, ||\boldsymbol{\chi}-\boldsymbol{A}(B)||_{L_2}^2+\lambda||B||_*
       \right).
    \end{aligned}
\end{equation}
Here, $||\cdot||_*$ represents the nuclear norm of the matrix. 

Using recent theoretical results on recovery of low-rank Toeplitz matrices \cite{huang2025low},  we show that  this problem can be solved accurately if $K\geq \Omega(s\log ^2M)$. 
\begin{theorem}
    \label{huanglow} \cite{huang2025low}
    Let $||\boldsymbol{\epsilon}||_2\leq \eta$. With probability exceeding $1-e^{-cK}$, the solution $B^*$ to Eq.~(\ref{prb-lrm}) satisfies
    \begin{equation}
        ||B^*-B(S)||_F\leq C\frac{\eta}{K^{1/2}},
    \end{equation}
    where this bound holds simultaneously for all symmetric Toeplitz matrices $B$ of rank at most $s$, provided that $K>Ls\log^2M$. Here, $||\cdot||_F$ denotes the Frobenius norm, and $c, C$ and $L$ are some numerical constants.
\end{theorem}

In realistic physical systems, the noise spectrum $S(\omega)$ is non-negative, and so is $\mathbf{S'}$. As such, we have
\begin{equation}
    ||B(S)||_* \propto ||\mathbf{S'}||_{L_1} \approx ||\mathbf{S}||_{L_1}
\end{equation}
The problem in Eq.~(\ref{prb-lrm}) can then be rewritten in the same form as in the previous section:
\begin{equation}
    \label{prb-lrmr}
        \begin{aligned}
\mathbf{S}^* 
    = \underset{\mathbf{S}: G_N \to \mathbb{R}}{\arg\min} 
       \left( \, \| \boldsymbol{\chi} - \mathbf{F}\mathbf{S} \|_{L_2}^2 
       + \lambda \, \|\mathbf{S}\|_{L_1} \, \right).
    \end{aligned}
\end{equation}
Since both $B(S)$ and $B^*$ are Toeplitz and diagonal in the Fourier basis, we have
\begin{equation}
\label{eqn-trans}
||B^*-B(S)||_F = M||\mathbf{S^*}\sinc^2(\tfrac{\omega\tau}{2})-\mathbf{S}\sinc^2(\tfrac{\omega\tau}{2})||_2.
\end{equation}
In practice, we can set the number of grid points $N=M$. Note that $\sinc^2(\tfrac{\omega\tau}{2}) $  varies
mildly between $\tfrac{4}{\pi^2}$ and $1$, when $\omega$ is in the interval $[-\tfrac{\pi}{\tau}, \tfrac{\pi}{\tau}]$. Thus, estimating the sparse spectrum of $S(\omega)$ is equivalent to estimation of $S(\omega)\sinc^2(\tfrac{\omega\tau}{2})$, with a small loss of precision up to some constant factor.  Substituting (\ref{eqn-trans}) into Theorem~\ref{huanglow}, we get that with probability exceeding $1-e^{-cK}$, the solution $\mathbf{S}^*$ to Eq.~(\ref{prb-lrmr}) satisfies
\begin{equation}
    ||\mathbf{S}^*-\mathbf{S}||_2\leq C\frac{\eta}{NK^{1/2}},
\end{equation}
provided that $K>Ls\log^2N$, and where $c, C$ and $L$ are some numerical constants.

As a result, despite using a different measurement setup (uncorrelated random pulses) and a different mathematical formulation (low-rank Toeplitz matrix recovery), we arrive at a place that is reassuringly familiar: we can still use the same CS algorithms as before to reconstruct sparse noise spectra efficiently. We call this CS method combined with Rademacher measurements CS$_\text{R}$.

\begin{figure}
\begin{centering}
\sidesubfloat[]{\label{2nd}%
  \includegraphics[width=0.8\textwidth]{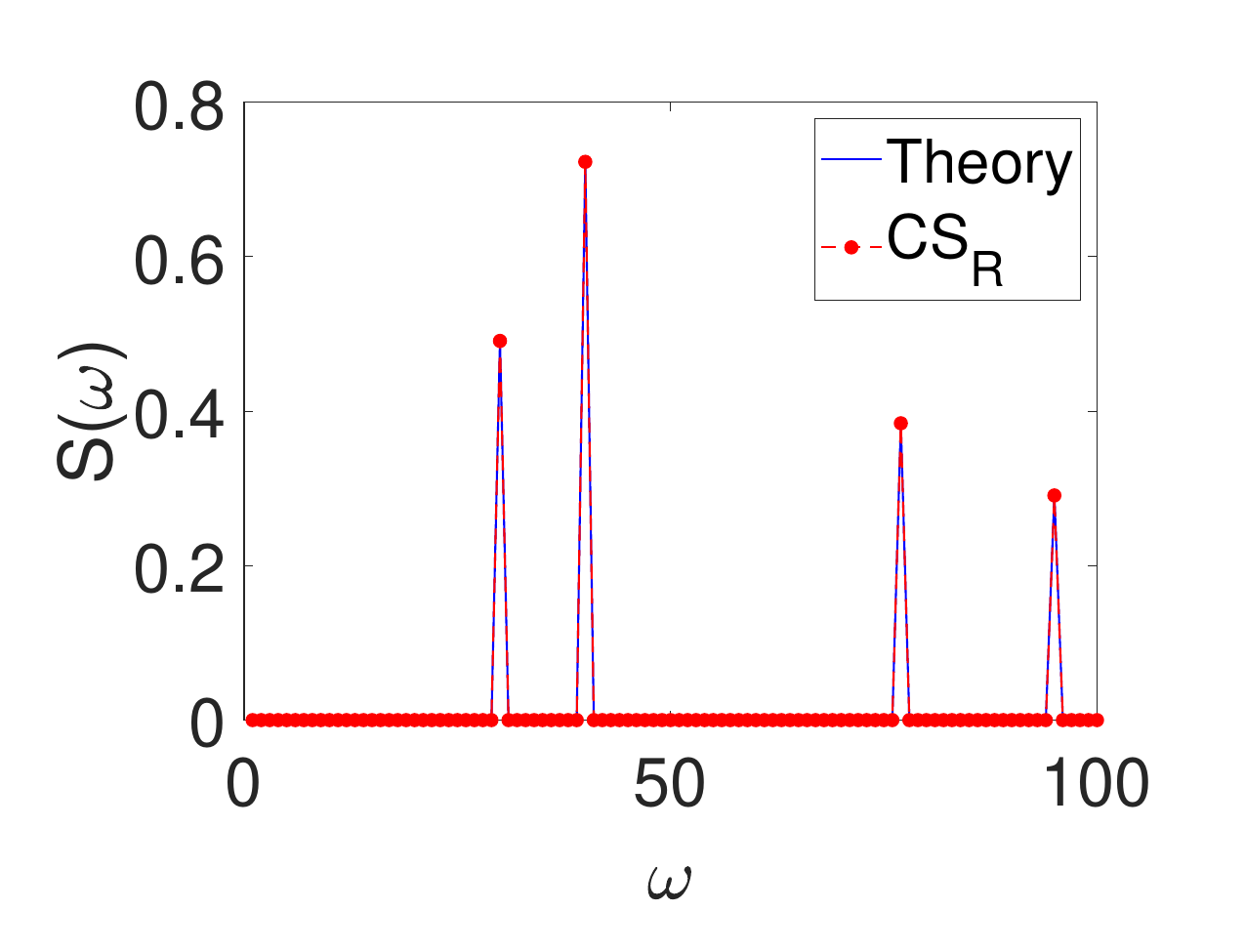}%
}
\par\end{centering}
\begin{centering}
\sidesubfloat[]{\label{unco_phase}%
  \includegraphics[width=0.8\textwidth]{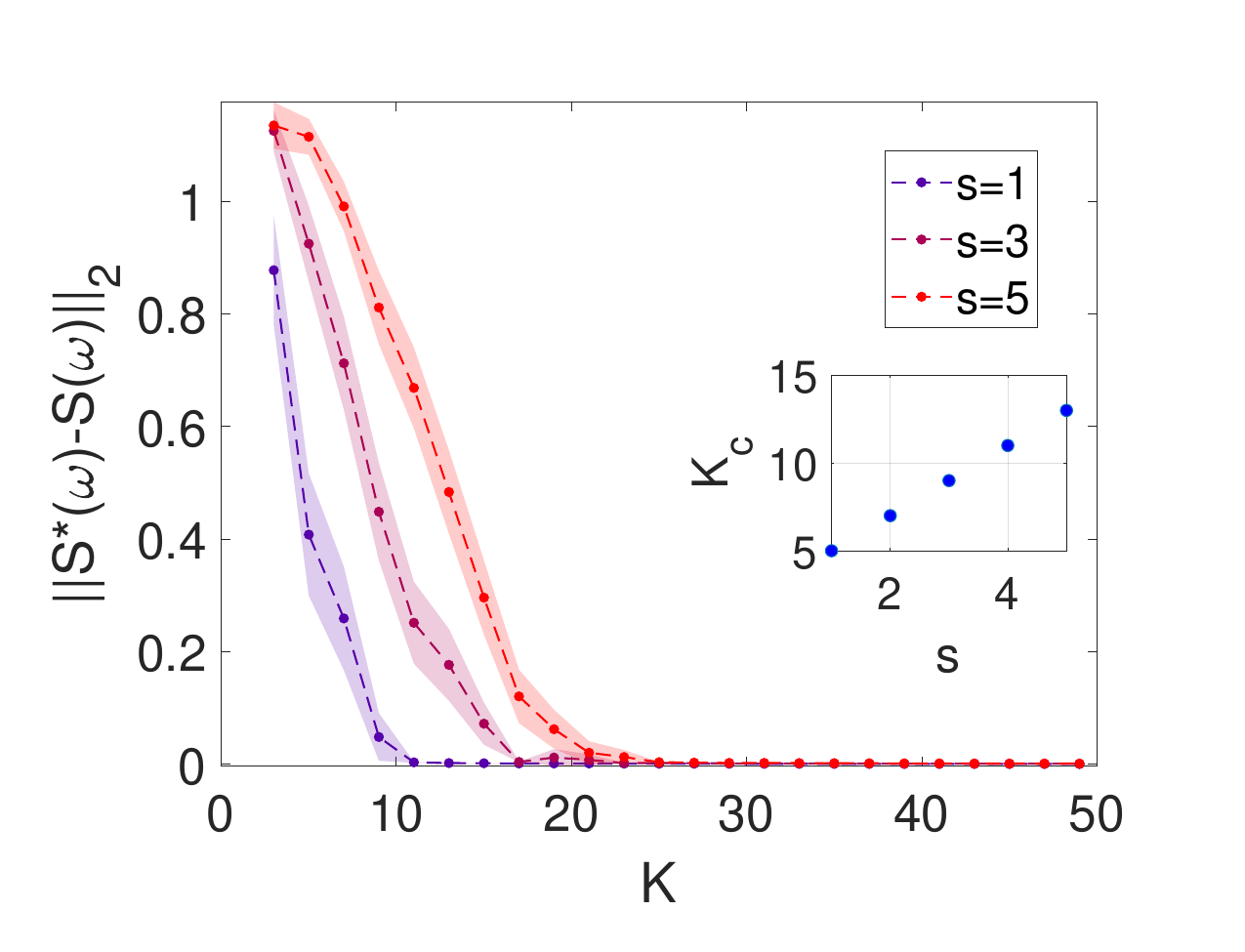}%
}
\par\end{centering}
\begin{centering}
\sidesubfloat[]{\label{unco_logn}%
  \includegraphics[width=0.8\textwidth]{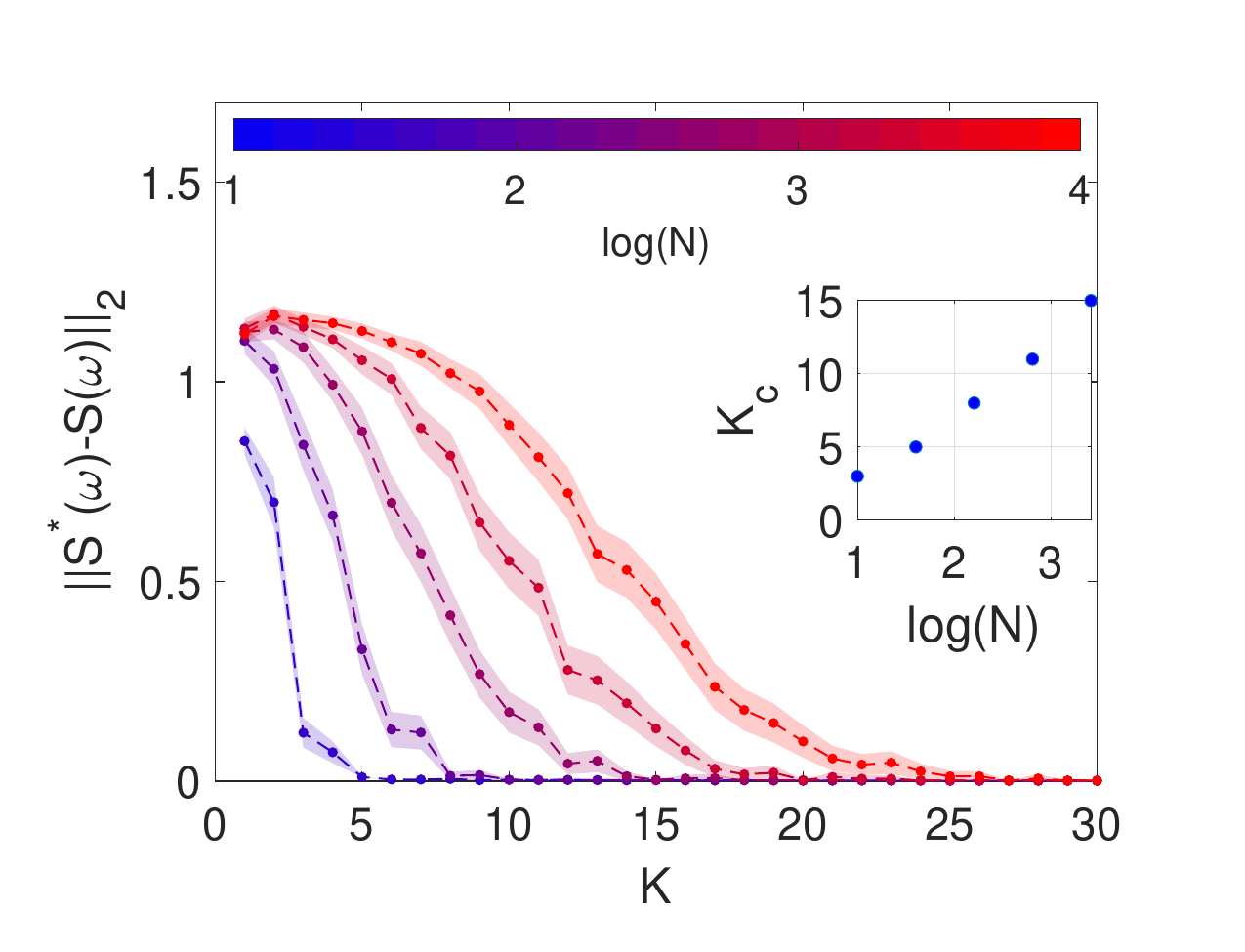}%
}
\par\end{centering}
\caption{
(a) A reconstruction of an ideal sparse spectrum using CS$_\text{R}$, the Rademacher-measurements-based CS method. The solid blue line represents a randomly generated spectrum with $N=100$ grid points. This spectrum is 4-sparse. The red line represent the decomposed spectrum using CS$_\text{R}$ based on $K=20$ different Rademacher sequences, with each repeated 5000 times. (b) The accuracy of CS$_\text{R}$ in reconstructing ideal sparse spectra as a function of $K$, the number of different Rademacher sequences. Different dotted lines represent different sparsities $s$, considering  100  randomly generated spectra with $N=100$. (c) The accuracy of CS$_\text{R}$ as a function of $K$, the number of different Rademacher sequences. Different dotted lines (different colors) represent different grid numbers, $N$, in a logarithmic scale. Each dot contains the simulation of 100 random spectra. Inset: The scaling of the critical number of Rademacher measuremetns, $K_c$, as a function of the logarithm of the grid numbers, $\log(N)$. In (b) and (c), each simulation is repeated 100 times and the shaded areas represent the 95\% confidence regime.}
\label{rpsillus}
\end{figure}

Fig.~\ref{2nd} presents a simulation that shows the efficiency of the Rademacher measurements technique. The solid blue line depicts the true spectrum, a 4-sparse signal defined over $N = 100$ grid points. Using only $K = 20$ random Rademacher measurements,  the CS$_\text{R}$ method accurately reconstructs the spectrum, shown as red circles. 

We then investigate the accuracy of this new method as a function of the number of Rademacher sequences, $K$, in Fig.~\ref{unco_phase}. For each value of the sparsity, \( s\), we generate 100 random spectra with a constant norm, each with \( N = 100 \) grid points, and compute the average reconstruction error. We again quantify the accuracy using the \( L_2 \) distance between the discrete true spectrum, \( S(\omega) \), and the CS$_\text{R}$ estimate \( S^*(\omega) \). Each dotted line in Fig.~\ref{unco_phase} exhibits a distinct phase transition: as \( K \) increases, the error drops sharply around a threshold value. We define this threshold \( K_c \) as the minimum number of basis functions needed for the reconstruction error to fall below 0.5. For example, \( K_c = 9 \) when \( s = 3 \). 

Fig.~\ref{unco_logn} is similar to the simulations in Fig.~\ref{unco_phase}, in which we show the accuracy of CS$_\text{R}$ in reconstructing ideal sparse spectra as functions of $K$. Different dotted lines (different colors) represent simulations with different grid numbers, $N$. For each value of $N$, we fix the sparsity $s=2$, and  generate 100 random spectra (again with fixed norm) to compute averaged results. The distinct phase transition of the accuracy is obtained as \(K\) increases. As shown in the inset of Fig.~\ref{unco_phase} and \ref{unco_logn}, \( K_c \) scales linearly with \( s \) and quadratically with \(\log N\), in agreement with the theoretical prediction that \( K \sim s^* \log^2 N \).

Overall, the above results using CS with Rademacher measurements are comparable to the previous results of \cite{huang2025random} using CS with Fourier basis measurements, for reconstructing sparse noise spectra. This shows that the advantages of Rademacher measurements over Fourier basis measurements (namely, simpler experimental implementation) can be obtained without paying a large price in the number of measurements or the reconstruction accuracy.

It is natural to ask whether Rademacher measurements can be combined with the techniques of Section \ref{piecewise-th} in order to reconstruct piecewise-linear noise spectra. While the answer to this question does not seem obvious due to the issues mentioned in Section \ref{piecewise-th}, we briefly explore this question in the next section, for a specific family of noise spectra that appear in quantum dot systems.

\section{\label{appli}Applications}
\subsection{\label{simu}Quantum Dots}
Next, to quantify the performance of our new methods for realistic physical systems, we examine their ability to reconstruct the spectral density of noise that occurs in InAs/GaAs quantum dots. That is, we perform qubit noise spectroscopy, where the role of the qubit is played by the electron spin, and the noise arises from hyperfine interactions between the electron spin and a surrounding ensemble of nuclear spins broadened by strain \cite{stockill2016quantum, farfurnik2021all}. 

\begin{figure}
\begin{centering}
\sidesubfloat[]{\label{four_qt}%
  \hspace{0.9cm}\includegraphics[width=0.65\textwidth]{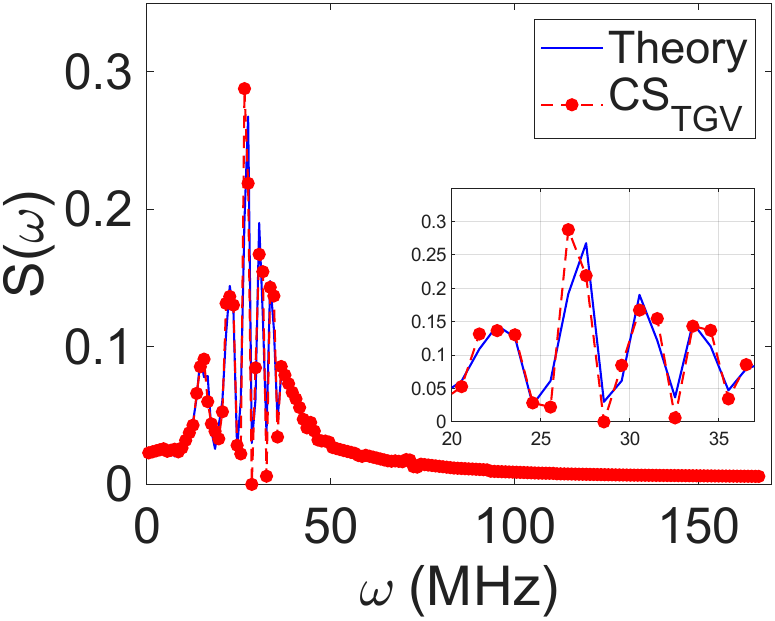}%
  \hspace{0.5cm}
}
\par\end{centering}
\vskip 11pt
\begin{centering}
\sidesubfloat[]{\label{unco_illus}%
  \hspace{0.9cm}\includegraphics[width=0.65\textwidth]{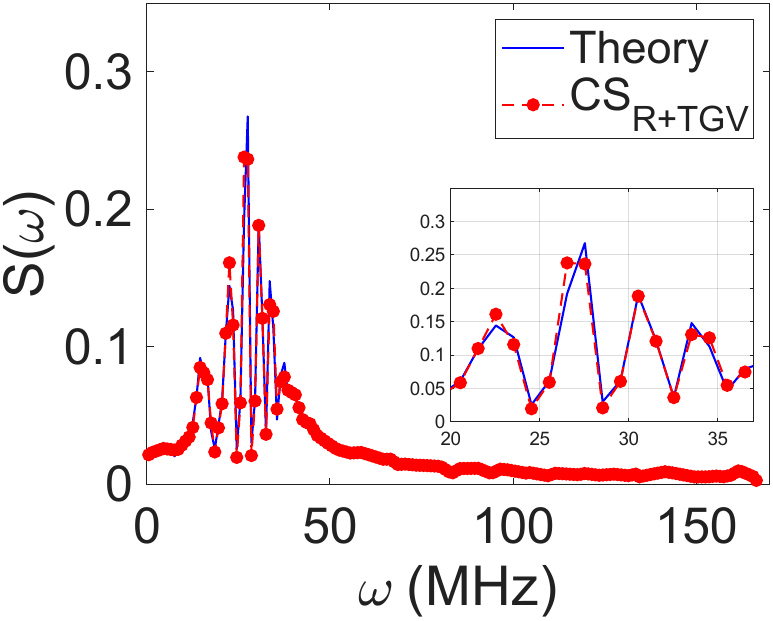}%
  \hspace{0.5cm}
}
\par\end{centering}
\begin{centering}
\sidesubfloat[]{\label{compare}%
  \includegraphics[width=0.8\textwidth]{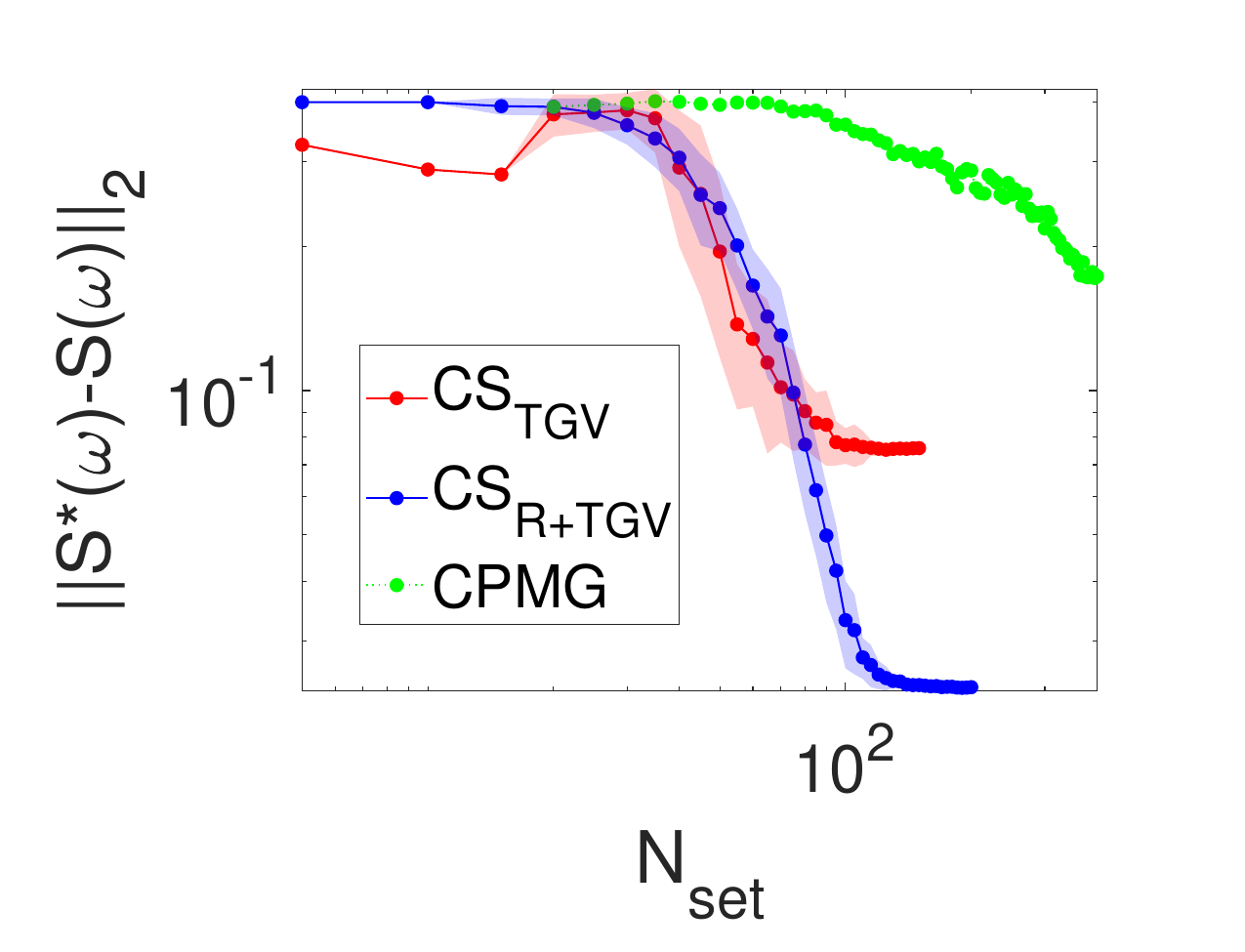}%
}
\par\end{centering}
\caption{(a) Reconstruction of the noise spectrum of an ensemble of nuclear spins interacting with an InAs/GaAs quantum dot (under an external magnetic field of B = 2 T at the Voigt geometry) using CS$_\text{TGV}$. The blue solid line represents the theoretically simulated noise spectrum, scaled to have maximum intensity 1, and then downsampled (via averaging) onto a grid containing 167 points in the frequency domain; this is comparable to the grid spacing used in experiments such as \cite{farfurnik2021all}. The red dotted line represents the simulated reconstructed spectrum on the same grid of 167 points, obtained via random pulse sequences with $(M, N_1, N_2)=(200, 200, 50)$ and $K=70$ different Fourier basis functions. (b) Same experiments with the Rademacher measurements method. The blue solid line represents the theoretically simulated noise spectrum. The red dots represent the simulated reconstructed spectrum considering random pulse sequences with $K=90$ different sequences. (c) Accuracy of reconstructing the InAs/GaAs noise spectrum as a function of the number of sets of experiments, $N_\text{set}$. The solid blue line and the dashed red line represent the accuracy of CS with second-order TGV,  and Rademacher measurements combined with TGV. The CS simulations are repeated for 30 times and the shaded areas represents 95\% confidence regimes. The dotted green line represents the reconstruction accuracy of the noise spectrum using CPMG sequences.}
\end{figure}

The solid blue line in Fig.~\ref{four_qt} shows the theoretical noise
spectral density. The spectral density was calculated as the Fourier transform of the autocorrelation function, $\langle V(t)V(0) \rangle_V$, where $V(t)$ represents a single realization of a  magnetic field at the quantum dot's location that takes into account an external magnetic field of $2$T applied perpendicular to the sample growth axis (Voigt geometry) as well as a fluctuating (Overhauser) field generated by the nuclei in the bath. The calculation of this autocorrelator requires summing over the contributions of $8,000$ In spins, $12,000$ Ga spins, and $20,000$ As nuclear spins with randomized orientations based on their Hamiltonian, which consists of a Zeeman splitting term under the external magnetic field as well as a quadrupolar coupling term due to inhomogeneous strain. 

The resulting simulated spectrum features several narrow resonances that appear around the 
Larmor precession frequencies of the In, As, Ga nuclei, with different resonant frequencies corresponding to different magnetic spin quantum numbers of the nuclei, $m_I$, and  linewidths of resonances dictated by the strain in the environment. Cooling the quantum dots to a temperature of $\sim 4 K$ ensures that longitudinal relaxation is negligible and the spin dynamics of the quantum dot electron spin (e.g., under the application of the CPMG or random pulse sequences) are fully dictated by the simulated noise spectral density. A full description of the theoretical model and simulation procedures can be found in the Supplementary Material of references \cite{stockill2016quantum, farfurnik2021all}.


First, we demonstrate the recovery of piecewise-linear approximations of this noise spectrum, using the technique from Section \ref{piecewise-th}. The red dotted line in Fig.~\ref{four_qt} shows the discretized spectrum reconstructed using the CS$_\text{TGV}$ method with \( K = 70 \) randomly chosen Fourier basis functions. The reconstruction successfully recovers both the positions and shapes of the narrow peaks, as well as the slowly decaying broadband background, demonstrating the method's effectiveness in capturing realistic spectral features with limited measurements. In contrast, the previous method only provides the positions of peaks.

Note that the noise spectrum of the previously introduced InAs/GaAs quantum dot system can also be approximated by a sparse function (albeit with poorer accuracy and sparsity, than the piecewise-linear approximation). So, we can again use it to test the new Rademacher random pulse sequences method of Section \ref{unco_th}. Inspired by the previous section, we modify the Lagrangian in Eq.~(\ref{prb-lrmr}) by adding a penalty term on the discrete Laplacian of $\mathbf{S}$:
\begin{equation}
\label{CS_un}
    \begin{aligned}
        &\mathbf{S^*} = \underset{\mathbf{S}: G_N \to \mathbb{R}}{\arg\min} 
        \left( ||\boldsymbol{\chi}-\mathbf{F}\mathbf{S}||_{L_2}^2 + \lambda_1||\mathbf{S}||_{L_1}+\lambda_2||D^2\mathbf{S}||_{L_1} \right).\\
    \end{aligned}
\end{equation}
Equation (\ref{CS_un}) is a combination of the TGV and Rademacher measurements (CS$_\text{R+TGV}$). As is shown in Sec.~\ref{piecewise-th}, the second order TGV penalty term restricts the recovered spectrum to be piecewise-linear, thus revealing more details than a simple sparse signal recovery. 

Fig.~\ref{unco_illus} shows the discretized spectrum (red dots) reconstructed by CS$_\text{R+TGV}$ from $K = 90$ randomly chosen Rademacher sequences. The method accurately recovers the narrow peaks and   decaying broadband background.

We further quantify the accuracy of extracting the InAs/GaAs noise spectrum under different approaches. The reconstruction accuracy is defined as the $L_2$ norm of the difference between the reconstructed spectrum, $S^*(\omega)$, and the true spectrum, $S(\omega)$ (discretized appropriately). The dotted red and blue lines in Fig. \ref{compare} represent the simulated reconstruction accuracies of CS$_\text{TGV}$ and CS$_\text{R+TGV}$, respectively. For both methods, we have $N=200$ grid points.  In these simulations, we assume no experimental errors  and only focus on the effect of the number of different sets of experiments, $N_\text{set}$. (Note that even when there are no experimental errors, the reconstruction of the noise spectrum is still expected to be imperfect, as we are approximating the true noise spectrum by a piecewise-linear function.) For both random pulse sequence methods, $N_{\text{set}}\approx K$.  And for both methods,  we observe a sharp change (phase transition) in the accuracy of reconstructing the spectrum. The CS$_\text{R+TGV}$ method seems to perform particularly well on this example, but we do not have a theoretical explanation for this, due to the issues mentioned in Section \ref{piecewise-th}.

We also compare the accuracies of both CS methods in resolving the InAs/GaAs noise spectrum to the ones obtained by using  conventional CPMG (dotted green line in Fig.~\ref{compare}). For CPMG, the number of sets of experiments required for noise spectroscopy is given by $N_\text{set}={2\omega_c T}/{\pi}$, i.e., the number of different sequences that probe the noise spectrum over the frequency range $[0, \omega_c]$ given the total experiment time, $T$. When $N_\text{set}< 100$,  CPMG cannot resolve the peaks because of poor resolution. When $N_\text{set} > 100$, the accuracy in reconstruction decreases with a constant rate. In contrast, both CS methods experience a sharp phase transition, and gain decent speedup over CPMG.

\subsection{\label{unco_p} Sensing with Fewer Pulses}

As discussed in Sec.~\ref{simu}, Rademacher random pulse sequences combined with TGV techniques can be used to 
reconstruct the noise spectra of a quantum dot system. However, in some experimental platforms, it is difficult to apply many pulses to the qubits, due to undesirable physical effects such as laser-induced tunneling, or practical limitations of the control electronics~\cite{farfurnik2021all}.

In this section, we show that we can even further reduce the number of pulses when performing Rademacher measurements. Recall from Sec.~\ref{unco_th} that the pulse sequence is controlled by a Rademacher random vector $\vec{U}$, with probability $p=0.5$ that $U_m = 1$, and probability $1-p = 0.5$ that $U_m = -1$.  The expected number of pulses, $N_p$, we need to apply in the Rademacher random pulse sequences equals
\begin{equation}
\label{equ_np}
    N_p = M2p(1-p).
\end{equation}
Eq.~(\ref{equ_np}) can be calculated by observing that we apply a pulse whenever $U_{m+1} \neq U_m$, which happens with probability $2p(1-p)$. As such, when $p$ deviates from $0.5$, the number of pulses decreases quadratically. We then investigate how the accuracy of the CS method changes when we adjust the value of $p$.

\begin{figure}
\includegraphics[width=0.8\textwidth]{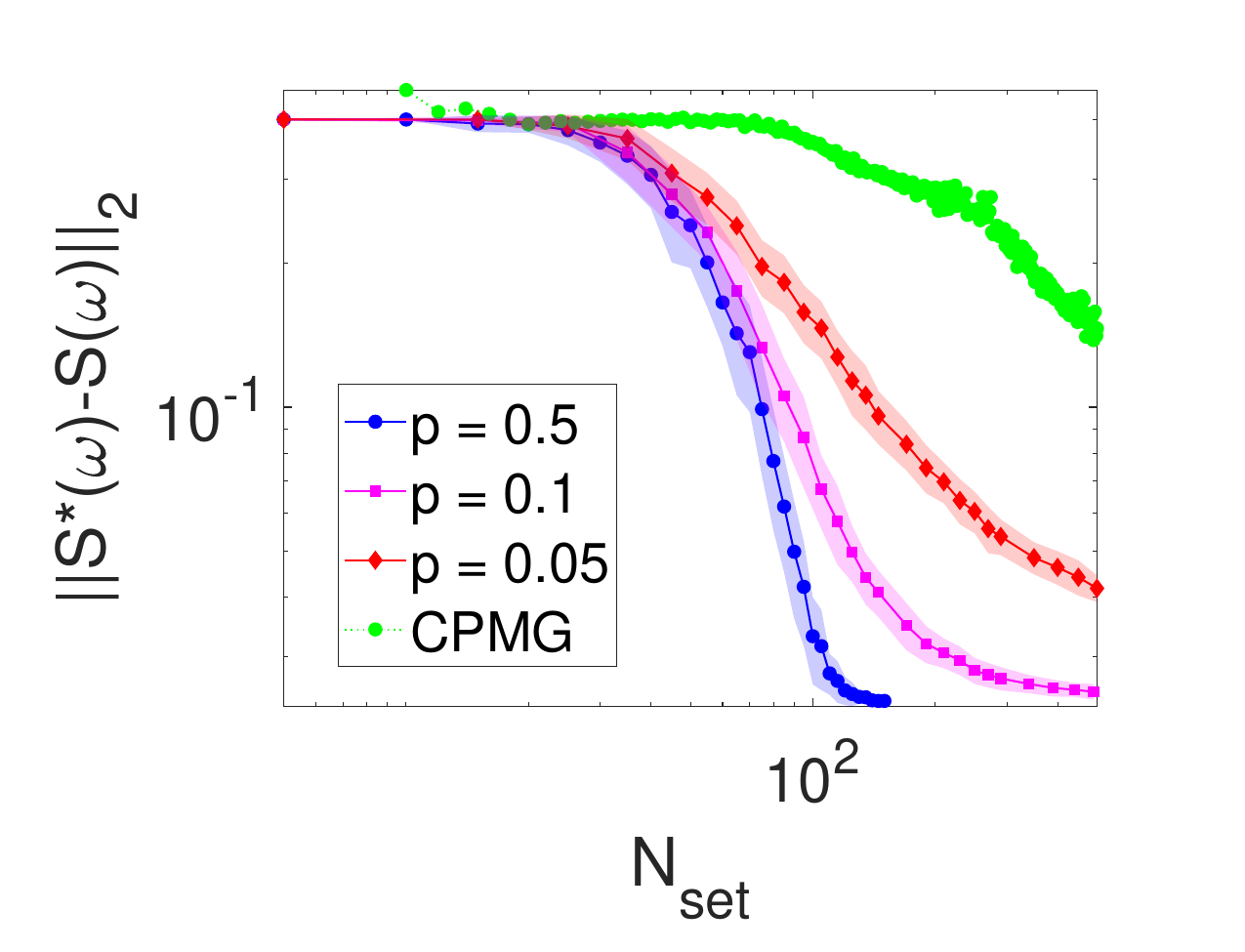}%
\caption{Simulation of the accuracy of reconstructing the InAs/GaAs noise spectrum as a function of the number of sets of experiments, $N_\text{set}$. The blue dotted, magenta squared and red diamond lines represent the accuracy of Rademacher pulse sequences with $M=N = 200$ and  $p = 0.5, 0.1, 0.05$, respectively. The corresponding averaged numbers of pulses are $N_p = 100, 36, 19$. The simulations are repeated 40 times and the shaded areas represents 95\% confidence regimes. The dotted green line represents the reconstruction accuracy of the noise spectrum using CPMG sequences. As $p$ decreases, the speedup for the CS method diminishes.}
\label{mixp}
\end{figure}

Fig.~\ref{mixp} shows the simulation of the  accuracy of reconstructing the InAs/GaAs noise spectrum, similar to what we have done in Fig.~\ref{compare}. The blue dotted, magenta squared, and red diamond lines represent results of the Rademacher measurements with probabilities, $p=0.5, 0.2,$ and $0.05$, respectively. In this simulation, the number of grid points equals the number of time segments, such that $N=M=200$. Corresponding expected values for the blue, magenta, and red dotted lines are calculated via Eq.~(\ref{equ_np}), with $N_p=100, 64$ and $19$. The green dotted line represents the result obtained from  CPMG. Note that the blue and green dotted lines are identical to the identically colored dotted lines in Fig.~\ref{compare}. 

We see that there is a regime for $p$ in which the number of pulses we need is greatly decreased, while the loss in the efficiency of reconstructing the spectrum of InAs/GaAs is bearable, making this a potentially useful trade-off. For example, by setting $p=0.1$, $N_p$ is decreased down from 100 to only 36. Meanwhile, the threshold number of $N_\text{set}$ for the accuracy to drop below 0.1 barely changed from 75 to 85, and remains an advantage over conventional CPMG.

\section{\label{out}Outlook}
To conclude, we improve on the random-pulse-sequence method for qubit noise spectroscopy in two respects. We expand the method's applicability to noise spectra with piecewise-linear features, via TGV regularization. We also simplify the implementation via Rademacher measurements, when reconstructing sparse noise spectra. The proposed methods are demonstrated using numerical simulations on realistic physical systems, such as optically-active quantum dots. Compared to previous work, these new developments broaden the reach of random pulse sequences and reduce the experimental complexity while preserving reconstruction accuracy. This brings the technique closer to experimental feasibility for quantum dots, and potentially also for other quantum systems, such as nitrogen vacancy centers \cite{romach2019measuring}.

For future research, we would like to further explore the class of noise spectra that can be rapidly reconstructed by random pulse sequences. For example, while piecewise linear modeling works well for many realistic physical systems, there are other possible approaches, such as model-based compressive sensing \cite{baraniuk2010model}, which may be useful for characterization of noise spectra in current quantum computing platforms. There is also room for improvement, as well as encouraging recent progress \cite{genzel2022compressed, huang2025low}, in the theoretical recovery guarantees for compressed sensing in the scenarios studied in this paper, i.e., for reconstruction of 1-dimensional signals, using TGV regularization, and Rademacher measurements.


\section{Acknowledgments}

This work was partially supported by the Institute for Robust Quantum Simulation (RQS) (NSF QLCI grant OMA-2120757). We thank Yu-Xin Wang for insightful discussions, and Joshua Bienfang and an anonymous referee for helpful comments on a previous draft of the paper. The data that support the findings of this article are available at \cite{huang_data}.
\bibliographystyle{ieeetr}
\bibliography{reference}

\end{document}